\documentclass[final]{svjour2}
\usepackage{graphicx}
\usepackage{rotating}
\usepackage{amssymb}
\usepackage{mathptmx}
\usepackage{bm}
\usepackage[version=3]{mhchem}
\usepackage{hyperref}
\usepackage[numbers,sort&compress]{natbib}
\usepackage[caption=false]{subfig}
\usepackage{microtype}
\usepackage{units}
\usepackage{upgreek}

\makeatletter
\journalname{Journal of Low Temperature Physics}

\bibpunct{}{}{,}{s}{}{,}

\begin{document}
\newcommand{\hdblarrow}{H\makebox[0.9ex][l]{$\downdownarrows$}-}
\title{Optical Response of Strained- and Unstrained-Silicon Cold-Electron Bolometers}
\author{T. L. R. Brien\textsuperscript{1} \and 
P. A. R. Ade\textsuperscript{1} \and 
P. S. Barry\textsuperscript{1} \and C. J. Dunscombe\textsuperscript{1} \and 
D. R. Leadley\textsuperscript{2} \and D. V. Morozov\textsuperscript{1} \and 
M. Myronov\textsuperscript{2} \and E. H. C. Parker\textsuperscript{2} \and 
M. J. Prest\textsuperscript{3} \and M. Prunnila\textsuperscript{4} \and 
R. V. Sudiwala\textsuperscript{1} \and T. E. Whall\textsuperscript{2} \and 
P. D. Mauskopf\textsuperscript{1,5}}
\institute{T. L. R. Brien \at tom.brien@astro.cf.ac.uk \\[1.5EX]
\textsuperscript{1} School of Physics and Astronomy, Cardiff University, Cardiff, CF24 3AA, United Kingdom\\%
\textsuperscript{2} Department of Physics, University of Warwick, Coventry, CV4 7AL, United Kingdom\\%
\textsuperscript{3} School of Engineering, Cardiff University, Cardiff, CF24 3AA, United Kingdom\\%
\textsuperscript{4} VTT Technical Research Centre of Finland, FI-02044 VTT Espoo, Finland\\%
\textsuperscript{5} Department of Physics and School of Earth \& Space Exploration, Arizona State University, 650 E. Tyler Mall, Tempe, AZ 85287, United States of America}
\maketitle

\begin{abstract}
We describe the optical characterisation of two silicon cold-electron bolometers each consisting of a small ($32 \times 14~\mathrm{\upmu m}$) island of degenerately-doped silicon with superconducting aluminium contacts. Radiation is coupled into the silicon absorber with a twin slot antenna designed to couple to 160-GHz radiation through a silicon lens. The first device has a highly-doped silicon absorber, the second has a highly-doped strained-silicon absorber. Using a novel method of cross-correlating the outputs from two parallel amplifiers we measure noise-equivalent powers of $3.0 \times 10^{-16}$  and $6.6 \times 10^{-17}~\mathrm{W\,Hz^{\nicefrac{-1}{2}}}$ for the control and strained device respectively when observing radiation from a 77-Kelvin source. In the case of the strained device, the noise-equivalent power is limited by the photon noise.  
\end{abstract}
\section{Introduction}
\par 
The cold-electron bolometer is a broadband bolometric detector where the only fundamental limit to the frequency response comes from the spectral transmission of the absorber. In order to tune the detector to respond to a certain frequency band, the absorber is coupled to an antenna structure which defines the frequency response.
The original, and still most common, cold-electron bolometer design involved a normal-metal absorber with tunnelling contacts to superconducting leads; see, for example, the work of Kuzmin.\cite{Kuzmin2001} This structure has also been used in the so-called \textit{microrefrigerator} devices to cool electrons to below the lattice temperature.
\par 
Savin et al.\cite{Savin2001} showed that if the normal-metal absorber is replaced by a degenerately doped, electrons can be more efficiently cooled due to the reduced electron-phonon coupling in such a material compared to a metal.\cite{Leoni1999} It has been shown that introducing strain into the silicon lattice can further reduce this coupling, allowing microrefrigerator-type devices to achieve lower electron temperatures compared to unstrained silicon.\cite{Prest2011}
\par
Based on this lower electron temperature achieved using strained silicon it is expected that the responsivity of cold-electron bolometer based on strained silicon should be higher than a comparable detector utilising unstrained silicon. A common advantage that both strained- and unstrained-silicon cold-electron bolometers share, compared to those using a normal-metal absorber, is that they do not require the fabrication of an insulating layer, since the naturally forming Schottky barrier may be used instead. Here we explore the performance of cold-electron bolometers utilising degenerately-doped silicon as the absorbing element. We take two devices: the \textit{control} device using doped silicon, and a device where the doped absorber is strained by a layer of \ce{Si_{0.8}}\ce{Ge_{0.2}}. We compare these devices by characterising them in a dark environment and also when illuminated. For both device the doping concentration was $N_{\mathrm{D}} = 4\times 10^{19}~\mathrm{cm^{-3}}$. In order to measure the noise of the devices themselves we present a novel concept of cross-correlating the outputs of two matched JFET amplifiers to remove the uncorrelated amplifier noise leaving only the device noise.
\section{Theory}
The cold-electron bolometer consists of an absorbing island with tunnelling contacts to superconducting leads either side of the absorber. In the absence of any bias, the Fermi energy is the same in the absorber and the superconducting contacts and thus carriers cannot tunnel between the two. However, in the presence of an external bias, applied across the structure, the Fermi level in the superconductor can be moved relative to that of the absorber, allowing tunnelling from the absorber to the vacant states above the superconductor's energy gap. This structure is described to greater detail by Gulobev and Kuzmin, and Brien.\cite{Golubev2001,Brien2014}
\par 
The current, $I$, flowing through each of the symmetric junctions is given by:
\begin{align}
I = \frac{1}{eR_{\mathrm{N}}}\int^{\infty}_{\varDelta} 
	&\frac{E}{\sqrt{E^{2}-\varDelta^{2}}} \times 
	\left[f\left(\!E-eV/2,T_{\mathrm{e}}\right) - f\left(\!E+eV/2,T_{\mathrm{e}}\right)\right]\,\mathrm{d}E\,,
	\label{eqn:IV}
\end{align}
where $R_{\mathrm{N}}$ is the normal-state resistance, due to tunnelling, of an individual tunnelling barrier, $\varDelta$ is half the superconducting bandgap, $V$ is the voltage across both junctions due to tunnelling (i.e. it does not include voltages due to the resistance of the absorber or any leakage across the junction), and $f\!\left(E,T\right)$ is the Fermi distribution of electrons at temperature $T$. This tunnelling current either increases or decreases the average energy of the electrons in the absorber.This change in energy can be expressed as a tunnelling power, $P$, given by:
\begin{align}
P = IV +\frac{2}{e^{2}R_{\mathrm{N}}}\int^{\infty}_{\varDelta}&\frac{E^{2}}
	{\sqrt{E^{2}-\varDelta^{2}}} \times 
		\left[f\!\left(E,T_{\mathrm{s}}\right)
	- f\!\left(-E,T_{\mathrm{s}}\right)\right.\nonumber \\
	&\left.- f\!\left(E+eV/2,T_{\mathrm{e}}\right)+
		f\!\left(-E+eV/2,T_{\mathrm{e}}\right)\right]\,\mathrm{d}E\,,
	\label{eqn:Ptun}
\end{align}
where $T_{\mathrm{e}}$ is the temperature of the electrons in the absorber and  $T_{\mathrm{s}}$ is the temperature of the normal-state electrons in the superconductor. A negative value of $P$ represents a reduction in the energy of the electrons in the absorber and thus corresponds to electron cooling. In a cold-electron bolometer it is this tunnelling power which cools the absorber. 
\par 
The temperature of the electrons in the absorber is also affected by the weak thermal link to the phonons. This thermal power, $P_{\mathrm{e\mbox{-}ph}}$, is given by:
\begin{align}
P_{\mathrm{e\mbox{-}ph}} = \varSigma\varOmega\left(T^{\beta}_{\mathrm{e}} - 
		T^{\beta}_{\mathrm{ph}}\right)\,, \label{eqn:Pe-ph}
\end{align}
where $\varOmega$ is the area of the absorber, $T_{\mathrm{ph}}$ is the phonon temperature, $\varSigma$ is a material constant that has been measured to be $5.2 \times 10^{8}$ and $2.0 \times 10^{7}~\mathrm{W\,K^{-6}\,m^{-3}}$ for the control and strained silicon materials,\cite{Prest2011} and the power $\beta$ have been found\cite{Prest2011} to be $6$.
\par 
The responsivity, when current biased, $S_{\mathrm{V}}$, of such a detector has been derived by Golubev and Kuzmin\cite{Golubev2001} and is given by:
\begin{align}
S_{\mathrm{V}} = \frac{\frac{-\frac{\partial I}
	{\partial T_{\mathrm{e}}}}{\frac{\partial I}{\partial V}}}
	{\beta\varSigma\varOmega T^{\beta-1}_{\mathrm{e}}
	+\frac{\frac{\partial I}{\partial T_{\mathrm{e}}}}
	{\frac{\partial I}{\partial V}}\frac{\partial P}{\partial V}
	-\frac{\partial P}{\partial T_{\mathrm{e}}}}\,. \label{eqn:responsivity}
\end{align}
\par 
The dark noise-equivalent power is comprised of several terms, corresponding to various thermal and electronic process, and has been derived, for a current-biased device, by Golubev and Kuzmin\cite{Golubev2001} to be:
\begin{align}
\mathit{NEP}^{2}_{\mathrm{CEB}} = 
	&\frac{\left<\partial V^{2}\right>_{\mathrm{amp}}}{S^{2}_{\mathrm{V}}}
	+ 2\beta k_{\mathrm{B}}\varSigma\varOmega
		\left(T^{\beta+1}_{\mathrm{e}}+T^{\beta+1}_{\mathrm{ph}}\right)\nonumber\\
	&+\left<\partial P^{2}\right>
	+ \frac{\left<\partial I^{2}\right>}
		{\left(\frac{\partial I}{\partial V}S_{\mathrm{V}}\right)^{2}}
	- 2\frac{\left<\partial P\, \partial I\right>}
		{\frac{\partial I}{\partial V}S_{\mathrm{V}}}\,, \label{eqn:darkNEP}
\end{align}
where $\left<\partial V\right>_{\mathrm{amp}}$ is the noise voltage introduced by the readout, $k_{\mathrm{B}}$ is Boltzmann's constant, $\left<\partial P\right>$ is the heat-flow noise, and $\left<\partial I\right>$ is the current-shot noise. In the presence of optical power, the noise-equivalent power can be further degraded by noise introduced by the absorption of photons. This photon-noise limit to the noise-equivalent power is given by the well-known equation:\cite{Zmuidzinas2003}
\begin{align}
\mathit{NEP}^{2}_{\mathrm{photon}} = 2h\nu P_{\mathrm{opt}} 
	+ \frac{P^{2}_{\mathrm{opt}}}{\Delta\nu}\,, \label{eqn:photonNEP}
\end{align}
where $h$ is Planck's constant, $\nu$ and $P_{\mathrm{opt}}$ are the frequency and power of the radiation incident upon the detector, and $\Delta\nu$ is the optical bandwidth.
\section{Cross-Correlated Readout}
\begin{figure}[htb]
\centering
\includegraphics[width = 0.7\textwidth]{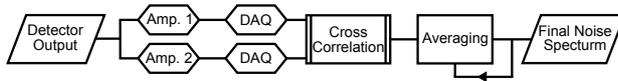}
\caption{Process flow for measurement of cross-correlated noise spectra in order to measure below the amplifier noise limit.}
\label{fig:crossCol_flow}
\end{figure}

As is often the case, the internal noise of device under test here was lower than the input referred noise of the amplifier used ($1~\mathrm{nV\,Hz^{\nicefrac{-1}{2}}}$ in this case). To combat this we have used a novel solution of feeding the output of the device to two matched JFET amplifiers. The output of these two amplifiers was then fed to a data-acquisition system and cross-correlated in software. This treatment has the effect of removing the amplifier, which is uncorrelated, and leaving only noise present in the system prior to the splitting of the signal wiring. This approach is illustrated in Fig.~\ref{fig:crossCol_flow}. Since the noise of each amplifier is random there is a finite chance of correlations between the two for any particular acquisition, due to this, in order to sufficiently reduce the amplifier noise in the final spectrum it is necessary to average multiple cross-correlations.
\par 
In reality there will always be a weak level of correlated amplifier noise in the final measurement due to stray capacitances in the system which provide a route for the amplifier noise to couple to the device. This low level of correlated amplifier noise presents itself as a noise floor below which further averaging offers little improvement. 
%
\section{Device Design}
Both the control and strained devices have been made and fabricated to a common design and process. For both devices the absorber is a $30~\mathrm{nm}$ tall mesa formed from etching n\textsuperscript{++}-doped silicon ($N_{\mathrm{D}} = 4 \times 10^{19}~\mathrm{cm^{-3}}$) and the contacts are sputtered aluminium. For the strained device the doped silicon is strained by a \ce{Si_{0.8}}\ce{Ge_{0.2}} buffer layer between the absorber and the silicon substrate. The sputtered aluminium is etched to form a twin-slot antenna, designed to respond to 160-GHz radiation.
%
\section{Experiment Setup}
Both devices have been tested in two scenarios. Firstly, dark (looking at a blank at $T_{\mathrm{ph}}$) tests were performed to verify the formation of a Schottky contact and to baseline device performance. The devices were then illuminated by radiation from external black-body sources (eccosorb) with temperatures of $77$ and $300~\mathrm{K}$. Radiation was coupled to the detector via a pair of back-to-back horns and a hemispherical silicon lens which focussed the radiation on to the twin-slot antenna. The power incident on the detector was limited by both a low-pass filter, with an edge frequency of $300~\mathrm{GHz}$, and by a $16~\mathrm{mm}$ limiting aperture. The detector was biased via a pair of $1~\mathrm{M\Omega}$ resistors and readout via cross-correlated JFET amplifiers (as discussed above). All optical measurements were performed at a bath temperature ($T_{\mathrm{bath}} = T_{\mathrm{ph}}$) of $350~\mathrm{mK}$. 
%
\section{Results}
\begin{figure}[htb]
\centering
\subfloat[]{
\includegraphics[width = 0.6\columnwidth]{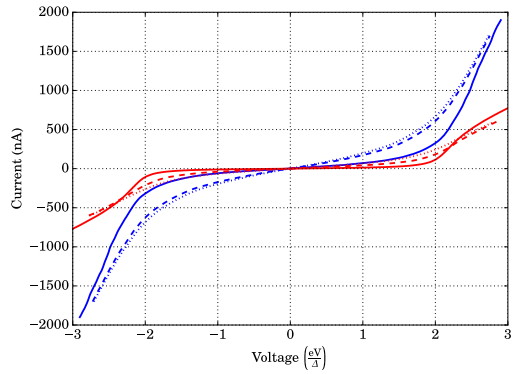}
\label{fig:IV}}\\
\subfloat[]{
\includegraphics[width = 0.6\columnwidth]{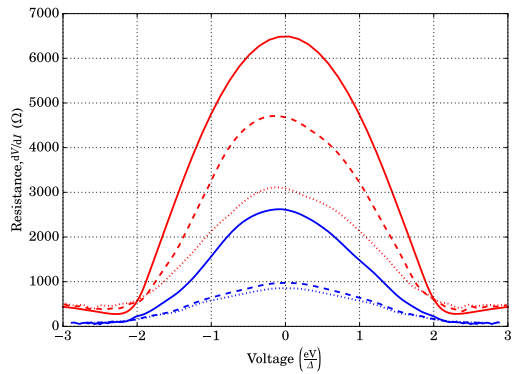}
\label{fig:Rderiv}}
\caption{Experimental results: (a) Current-voltage characteristics for control (blue, inside pair) and strained (red, outside pair) silicon cold-electron bolometers, measured in the presence of incident radiation from a 77-Kelvin (dashed lines) and 300-Kelvin (dotted lines) black-body source and when operated in darkness (solid line). (b) Differential resistance of the structures when operated in darkness and observing 77- and 300-K sources, colours and lines as in (a). (Colour figure online)}
\label{fig:data}
\end{figure}
To compare the strained- and unstrained-silicon cold-electron bolometers, both current-voltage ($I$-$V$) characteristics as well as noise spectra of the devices have been measured, these measurements have been performed dark and in the presence of optical power. $I$-$V$ curves are presented in Fig.~\ref{fig:IV}. For both devices there is a clear voltage response to increasing the source temperature from $77~\mathrm{K}$ (dashed lines) to $300~\mathrm{K}$ (dotted lines) and this can be seen to a greater extent when comparing these curves to the dark data (solid lines). From Fig.~\ref{fig:Rderiv} it can be also seen that the unstrained device (blue curves) exhibits a lower normal-state ($eV/\varDelta > 2$) resistance along with greater sub-gap leakage. This is attributed to lower-quality Schottky contacts in this device along with the absorber resistance being lower in this material ($50~\mathrm{\Omega}$ compared to $75~\mathrm{\Omega}$ in the strained detector).
\par 
\begin{figure}[htb]
\centering
\subfloat[]{
\includegraphics[width = 0.6\columnwidth]{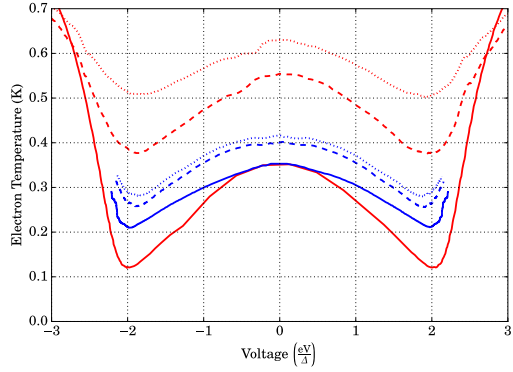}
\label{fig:Te}}\\
\subfloat[]{
\includegraphics[width = 0.6\columnwidth]{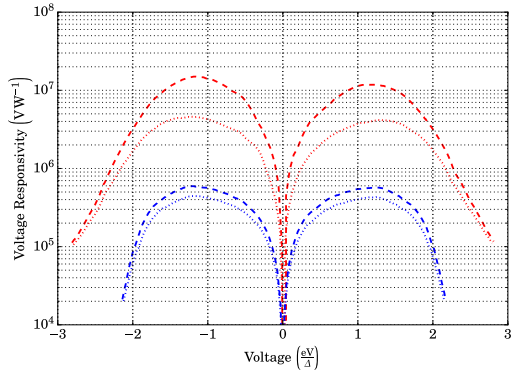}
\label{fig:responsivity}}
\caption{Calculated quantities: (a) Electron temperature computed from fitting the data shown in Fig.~\ref{fig:IV} to Eq.~\ref{eqn:IV}, (b) Voltage responsivity for the strained detectors. Colours and lines as in Fig.~\ref{fig:data}. In these plots the voltage has been corrected for any parasitic resistance, leaving only the tunnelling resistance, as required for Eq.~\ref{eqn:IV}. (Colour figure online)}
\end{figure}
\begin{figure}[htb]
\centering
\includegraphics[width = 0.6\columnwidth]{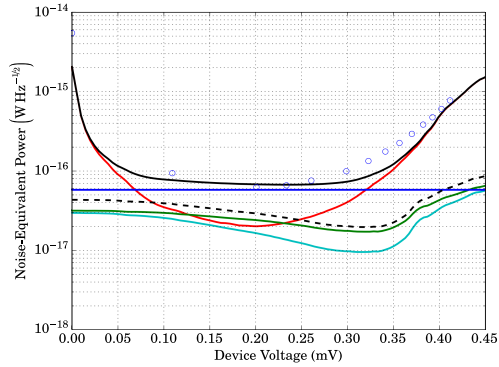}
\caption{Example of measured and modelled noise for strained detector observing a 77-Kelvin source. Noise modelled using Eq.~\ref{eqn:darkNEP} and Eq.~\ref{eqn:photonNEP}. Lines: modelled data, red - amplifier noise, green - tunnelling noise, cyan - e-ph noise, blue - photon noise, dashed black - total device noise (sum of tunnelling and e-ph noise), black - total modelled noise level. Circles: measured data. Overall the model and data are in good agreement and show that the device is photon noise limited from approximately $0.1$ to $0.3~\mathrm{V}$. (Colour figure online)
\label{fig:noise}}
\end{figure}

The electron temperature for these data has been computed by fitting the data to Eq.~\ref{eqn:IV}. Using the electron temperature at zero bias, along with Eq.~\ref{eqn:Pe-ph}, it is possible to calculate the power absorbed within the silicon in each scenario. For the control device the absorbed power was $34~\mathrm{pW}$ for the 77-Kelvin source and $43~\mathrm{pW}$ for the 300-Kelvin source. For the strained detector the absorbed power was $9.2~\mathrm{pW}$ for the 77-Kelvin source and $20~\mathrm{pW}$ for the 300-Kelvin source. The difference between these powers for the two detectors is due to alterations in the optical setup and in both measurements we believe there to be a substantial background power due to the cryostat optics and stray light; these values for the absorbed power are supported by our noise modelling, an example of which is shown in Fig.~\ref{fig:noise}. From the computed electron temperature it is also possible to calculate the responsivity using Eq.~\ref{eqn:responsivity}, the results of this are shown in Fig.~\ref{fig:responsivity}.
\par
Fig.~\ref{fig:responsivity} shows that when the incident power is increased there is an overall drop in the responsivity of the detector (the difference between the dashed and dotted lines), this is attributed to an overall increase in the electron temperature for the higher incident power. If the two curves for the 77-Kelvin source are examined (dashed lines in Fig.~\ref{fig:responsivity}), then the difference in peak responsivity is a factor of $\sim 25$. This value is very close to the difference in the $\varSigma$ parameter for the two materials ($\varSigma_{\mathrm{control}}/\varSigma_{\mathrm{strained}} = 26$).
\par 
From the responsivity, and by using Eq.~\ref{eqn:darkNEP} and Eq.~\ref{eqn:photonNEP}, it is possible to calculate the noise-equivalent power for the two detectors. This has been found to be $3.0 \times 10^{-16}~\mathrm{W\,Hz^{\nicefrac{-1}{2}}}$ for the control device and $6.6\times 10^{-17}~\mathrm{W\,Hz^{\nicefrac{-1}{2}}}$ for the strained device, both of these values are found when observing the 77-Kelvin source. In the case of the control device the limiting noise source, across the entire measurement, was the amplifier noise remaining after cross correlation (increased by the poorer responsivity). For the strained device the noise-equivalent power was photon-noise limited for biases corresponding to voltages in the range $\varDelta < V < 2\varDelta$ (as shown, for the 77-Kelvin source, in Fig.~\ref{fig:noise}), outside of this range the noise-equivalent power was limited by a combination of tunnelling and the amplifier noise.
\section{Conclusion}
The introduction of a straining layer to a silicon cold-electron bolometer has increased the detector responsivity by a factor of $25$. This increased responsivity can be attributed to the weaker coupling between the electrons and the phonons in the strained material. In terms of sensitivity the limiting noise source for the unstrained device was the readout, whereas in the strained detector the sensitivity was photon noise limited, afforded by the improved responsivity lowering the amplifier noise contribution. Taking the strained-silicon device we achieve a minimum noise-equivalent power of $6.6 \times 10^{-17}~\mathrm{W\,Hz^{\nicefrac{-1}{2}}}$ at optimum bias.
\begin{acknowledgements}
This work has been financially supported by the STFC through Grant ST/K000926/1, the EPSRC through grant numbers EP/F040784/1 and EP/J001074/1, and the Academy of Finland through Grant 252598.
\\
Information on how to access all data supporting the results in this article can found at Cardiff University data catalogue at doi: \href{http://dx.doi.org/10.17035/d.2016.0008254680}{10.17035/d.2016.0008254680}.
\end{acknowledgements}

\end{document}